\documentclass[aps,pra,twocolumn,showpacs,superscriptaddress,amsmath,floatfix]{revtex4-1}
\usepackage[dvipdf]{graphicx}
\usepackage{hyperref}
\usepackage{amssymb}
\usepackage{amsmath}
\usepackage{longtable}
\usepackage{rotating}
\usepackage{color}

\usepackage[T1]{fontenc}

\usepackage{fancyhdr}

\def\lesssim{\mathrel{\hbox{\rlap{\hbox{\lower4pt\hbox{$\sim$}}}\hbox{$<$}}}}
\def\gtrsim{\mathrel{\hbox{\rlap{\hbox{\lower4pt\hbox{$\sim$}}}\hbox{$>$}}}}
\def\cos{\rm cos}
\def\sin{\rm sin}
\newcommand{\be}{\begin{equation}}
\newcommand{\ee}{\end{equation}}
\newcommand{\bea}{\begin{eqnarray}}
\newcommand{\eea}{\end{eqnarray}}
\newcommand{\bdm}{\begin{displaymath}}
\newcommand{\edm}{\end{displaymath}}

\newcommand{\bg}[1]{\boldsymbol{#1}}
	
\newcommand{\mnras}{Monthly Notices of the Royal Astronomical Society}

\begin{document}
\title{Method to detect gravitational waves from an ensemble of known pulsars}
\author{Xilong Fan }
\address{Hubei University of Education, Wuhan, 430205, Hubei, China}
\address{SUPA, School of Physics and Astronomy, University of Glasgow, Glasgow, G12 8QQ, United Kingdom}
\author{ Yanbei Chen}
\address{Theoretical Astrophysics 350-17, California Institute of Technology, Pasadena, CA 91125, USA }
\author{ Christopher Messenger}
\address{SUPA, School of Physics and Astronomy, University of Glasgow, Glasgow, G12 8QQ, United Kingdom}

\begin{abstract}
Combining information from weak sources, such as known pulsars, for
gravitational wave detection, is an attractive approach to improve detection
efficiency. We propose an optimal statistic for a general ensemble of signals
and apply it to an ensemble of known pulsars. Our method combines $\mathcal
F$-statistic values from individual pulsars using weights proportional to each
pulsar's expected optimal signal-to-noise ratio to improve the detection
efficiency. We also point out that to detect at least one pulsar within an
ensemble, different thresholds should be designed for each source based on the
expected signal strength. The performance of our proposed detection statistic
is demonstrated using simulated sources, with the assumption that all pulsar
ellipticities belong to a common (yet unknown) distribution. Comparing with an
equal-weight strategy and with individual source approaches, we show that the
weighted combination of all known pulsars, where weights are assigned based on
the pulsars' known information, such as sky location, frequency and distance,
as well as the detector sensitivity, always provides a more sensitive detection
statistic. 
\end{abstract}

\maketitle

\section{Introduction}
Pulsars are believed to be rapidly rotating neutron stars (NSs) that can emit
continuous gravitational wave (GW) radiation if their mass distributions are
asymmetric~\cite{1979PhRvD..20..351Z}. Observations from first-generation GW
detectors have placed upper limits on the amplitude of these GWs from the known
galactic millisecond pulsars. This in turn allows constraints to be placed on
the ellipticities of these NSs~\citep{2010ApJ...713..671A}. With the advanced
detector era having recently begun with Advanced
LIGO~\cite{2015CQGra..32g4001L} in operation and Advanced
Virgo~\cite{2015CQGra..32b4001A}, and KAGRA~\cite{Aso:2013} close behind, we
will soon be able to make observations of these sources with significantly
increased sensitivity.


For each pulsar with known sky location and assumed GW phasing (as inferred
from arrival times of its radio pulses), time and frequency-domain
matched-filtering
approaches~\cite{2005PhRvD..72j2002D,Pitkin:2011cl,1998PhRvD..58f3001J,2014CQGra..31p5014A,2014ApJ...785..119A}
are commonly applied. The former has been used within the LIGO-Virgo
Collaboration for the known pulsar searches and applies a Bayesian
marginalization strategy to the unknown system
parameters~\cite{2005PhRvD..72j2002D}. The latter, frequency-domain approach,
known as the $\mathcal{F}$-statistic ~\cite{1998PhRvD..58f3001J} performs an
analytical maximization of the likelihood over the unknown parameters of each
pulsar and it is this method that we make use of for the remainder of this
paper. 

Combining sources to improve detection probability is an attractive
approach to weak signal detection (e.g.~detecting NS ellipticity from
analysis of the GW stochastic background~\cite{2014PhRvD..89l3008T} and
detecting gravitational wave memory using binary black hole
mergers~\cite{2016arXiv160501415L}). Since GW detectors currently study
${\sim}200$ known pulsars, the existing detection strategy for this relatively
large ensemble can be viewed as trying to detect each one separately, and then
waiting for the first detection to appear. This is certainly the most obvious
strategy to take, but not obviously the most optimal. Cutler and Schutz (CS)~\cite{2005PhRvD..72f3006C} proposed an alternative: first sum the
$\mathcal{F}$ statistic from each pulsar, and then use that sum as a new
detection statistic. In this initial study, CS used an equal weight for all the
pulsars to be combined.  One issue with this approach is that including pulsars
which are likely to emit relatively weak GWs  decreases the signal-to-noise
ratio (SNR) of the combined statistic. As indicated in their paper, the SNR of
the combined statistic  decreases  if the detection ensemble includes weak
sources where the squared SNR is less than half of the average squared SNR for
all observed pulsars. Therefore, to more efficiently detect GWs from an
ensemble of all known pulsars, it seems sensible to investigate the effects of
giving nonequal weights to the pulsars within the ensemble.

In this paper, we   generalize the idea proposed by CS, by considering the
prior distribution of GW strengths from the pulsars within the ensemble. After
a brief introduction to pulsar GW emission and the $\mathcal{F}$-statistic, we
  apply the general theory of hypothesis testing, and obtain a
Neyman-Pearson criterion for detecting GWs from an ensemble of pulsars. This
leads to an optimal detection statistic, which in idealized situations (i.e.,
when our prior knowledge of the signal and our model for the noise are an
accurate representation of reality) provides the highest detection probability
with a given false-alarm probability. As we  show, this statistic can in
some cases be approximated by linearly combining $\mathcal{F}$-statistic values
from the ensemble of pulsars with appropriate weights. 

We assume that the ellipticities of pulsars follow a common (yet unknown)
intrinsic distribution and that the orientation of their rotation axes is
isotropically distributed. We then draw on our knowledge of their sky location,
distance from the Earth, and their rotation frequency to construct prior
distributions on the expected GW amplitudes from our known pulsars. Since the
intrinsic ellipticity distribution remains unknown, we model it as a simple
exponential distribution, but perform tests using both exponential and Gaussian
distributions.


This paper is organized as follows. In Sec.~\ref{review}, we briefly review
the form of GW emission from individual pulsars and the
$\mathcal{F}$ statistic; in Sec.~\ref{f_statistic}, we introduce the optimal
statistic for a general ensemble of pulsars and discuss how it may apply to a
set of pulsars in idealized situations; in Sec.~\ref{sim}, we test our
statistic on two possible intrinsic distributions of pulsar ellipticity. We
summarize our main conclusions in Sec.~\ref{res}.

\section{Brief review of GW from known pulsars and the $\mathcal{F}$ statistic}
\label{review}

In this section, we give a brief overview of the signal model and
maximum-likelihood detection statistic for a single pulsar.

\subsection{Gravitational waveform}

For a single GW detector, the signal strain as a function of time, $h(t)$, from

\begin{align}
h(t) =\frac{16\pi^2\epsilon I f^2}{d}
\left[
\alpha_+ \tilde F_+(t) 
+\alpha_\times \tilde F_\times(t)\right] {\cos}[\Phi(t)+\Phi_0]\,,
\end{align}
with
\begin{align}
\alpha_+&= \frac{1+\cos^2\iota}{2} \cos 2\psi +
\cos\iota\,\sin2\psi  \,,\\
\alpha_\times&=-\frac{1+\cos^2\iota}{2} \sin 2\psi +
\cos\iota\,\cos 2\psi. 
\end{align}
Here we have assumed the pulsar, at distance $d$ from the Earth, to be an
triaxial ellipsoid rotating at frequency $f$ around one of its minor axes,
which stays constant in orientation. In~\cite{1998PhRvD..58f3001J} this is the
case when the angle between the total angular momentum vector of the star and
the star's axis of symmetry is $\pi/2$.

The pulsar is nearly spherical, with a moment of inertial $I$ around its
rotation axes, $\epsilon$ is its ellipticity, given by 
\begin{equation}
\epsilon=\frac{I_1-I_2}{I}
\end{equation}
with $I_1$ and $I_2$ being the two moments of inertia around the two principal axes
that are orthogonal to the rotation axis. The above four quantities
$(d,f,\epsilon,I)$ define the strength of the source as received at the
detector.    
 
In addition, $\tilde F_{+,\times}(t)$ are the (time-dependent, due to Earth's 
rotation) antenna patterns of the detector toward a source at the sky location
of the pulsar, while $\Phi(t)$ defines the GW phase evolution inferred from its
radio (or x-ray) pulsations. -both are considered known.  For the type of
emission we are considering, GW radiation will be emitted at twice the rotation
frequency, $2f$, with additional modulations due to the orbital motion of the
pulsar and the motion of the detector due to the Earth's rotation and orbit.
  
Finally, we have the polarization angle $\psi$ , the inclination angle $\iota$
that describes the pulsar's orientation, and $\Phi_0$ an additional unknown GW
reference phase, all of which we consider as unknown.  
  
In terms of notation, our $\iota$ and $\psi$ are the same as used
in~\cite{1998PhRvD..58f3001J}, while $\tilde F_+(t)$ and $\tilde F_\times(t)$
are respectively equivalent to $a(t)$ and $b(t)$ of~\cite{1998PhRvD..58f3001J}
where we have assumed that the angle between the two interferometer arms equals
$\pi/2$.

\subsection{The single-pulsar $\mathcal{F}$ statistic}
\label{sec:fstat}

Under the assumption that the measured strain is a combination of a GW signal
and additive detector noise $n$, with a single-sided noise special density
$S_{h}(f)$, the "near optimal" statistic is given by the so-called
$\mathcal{F}$ statistic, derived by Jaranowski, Krolak and
Schutz~\cite{1998PhRvD..58f3001J}. For point hypotheses with no uncertain model
parameters the maximum-likelihood approach of the $\mathcal{F}$-statistic is
optimal in the Neyman-Pearson sense whereby the detection probability $P_{\rm
DE}$ is maximized at fixed false-alarm probability $P_{\rm FA}$. However, even
for individual pulsar detection the signal model does include additional
unknown model parameters in which case the truly optimal approach is Bayesian
and requires marginalisation over those parameters~\cite{2008arXiv0804.1161S}.
Our investigation makes use of the $\mathcal{F}$-statistic as our input data
and hence by association also suffers from a lack of total optimality. However,
as shown in~\cite{2009CQGra..26t4013P} the reduction in sensitivity of the
$\mathcal{F}$-statistic over the fully optimal approach is slight.

For an observation time $T_{\rm obs}$, the $\mathcal{F}$-statistic satisfies a
$\chi^2$ distribution with 4 degrees of freedom (4-D) and has a noncentrality
parameter equal to the squared optimal SNR $\rho^2$, defined by 
\begin{equation}
\label{eqrho}
\rho^2 =\frac{256\pi^4 \epsilon^2 I^2 f^4 \mathcal{K} }{  d^2}\frac{T_{\rm obs}}{S_h(2f)} 
\end{equation} 
(note that $2f$ is approximately the gravitational wave frequency) with 
\begin{align}
\mathcal{K}&=\sum_{p,q=+,\times}\alpha_p\alpha_q F_{pq}\,,\nonumber\\
 F_{pq} & =\frac{1}{T_{\rm obs}}\int_{0}^{T_{\rm obs}}\tilde F_p(t) \tilde F_q(t) dt\,.
\end{align}
The unknown quantities defining the optimal SNR are the ellipticity $\epsilon$
and the geometrical factors contained within $\alpha_{+,\times}$ describing
the GW polarization and orientation of the pulsar. Note that $F_{++}\neq
F_{\times\times}$ and that averaging over many sidereal days leads to
$F_{+\times} \rightarrow 0$ and so such terms can be ignored.

For $\alpha_{+,\times}$ we shall assume that $\iota$ and $\psi$ are distributed
according to a random orientation of the pulsar's rotation axis. In this case,
points with coordinates $(\alpha_+,\alpha_\times)$ are distributed on the two-dimensional
plane axisymmetrically around the origin, with modulus 

\begin{equation}
\zeta \equiv \alpha_+^2+\alpha_\times^2  = \frac{1+6\,\cos^2\iota+\cos^4\iota}{4}
\end{equation}
and $\cos\,\iota$ uniformly distributed between $-1$ and $+1$. We can write
\begin{align}
\label{eqK}
\mathcal{K} =\zeta
\bigg[
F_{++}{\cos}^2(2\tilde\psi) +
F_{\times\times}{\sin}^2(2\tilde\psi) 
\bigg]
\end{align}
with $\tilde\psi$ related to $\psi$ by an offset,
\begin{equation}
4\tilde\psi = 4\psi - \arctan \frac{4\cos\iota(1+\cos^2\iota)}{\sin^4\iota}
\end{equation}
hence uniformly distributed between 0 and $2\pi$.  In this paper, we 
simply generate an ensemble of binaries using uniformly distributed
$\cos\,\iota$ and uniformly distributed $\psi$.  The average of $\mathcal{K}$
over this ensemble is given by
\begin{equation}
\langle \mathcal{K}  \rangle = \frac{\langle \zeta \rangle}{2} (F_{++} +F_{\times\times})  =\frac{2}{5}(F_{++} +F_{\times\times})
\end{equation}
It was shown by CS that for the detection of a single pulsar in a network of
$M$ detectors, the $\mathcal{F}$-statistic still satisfies  a 4-D $\chi^2$
distribution with a noncentrality parameter $\rho_{\text{net}}^2 = \sum_i^M
\rho_i^2$, where $\rho_i^2$ is the optimal single detector SNR as defined in
Eq.~\ref{eqrho}.

\subsection{Scaling of detectability with observation time}
\label{subsecthreshold}

In our idealized treatment with Gaussian noise,  the significance of
detection only depends on the noncentrality parameter $\rho^2$, which is
proportional to the observation time  $T_{\rm obs}$.  For this reason, the $T_{\rm
obs}$ required for a detection with a particular confidence level is
inversely proportional to $\mathcal{K}$ and $\epsilon^2$, or
\begin{equation}
T_{\rm det} =  \frac{S_h(2f) d^2}{256\pi^4\epsilon^2 I^2 f^4 \mathcal{K}}\rho_*^2
\end{equation} 
with $\rho_*$ being a threshold (or a sensitivity level for $\rho$)
determined by the desired false-alarm probability ($P_{\rm FA}$) and detection
probability ($P_{\rm DE}$), as we  discuss below.

Let us follow a frequentist approach of hypothesis testing. Suppose $X$ is
our detection statistic, which is either a 4-D $\chi^2$ distribution or a 4-D
noncentral $\chi^2$ distribution with noncentrality parameter $\rho^2$. Let
us first impose a detection threshold $X_{\rm th}$ on $X$, so that $P[X>X_{\rm
th}|\rho^2=0]=P_{\rm FA}$, which leads to 
\begin{equation}
\left(1+\frac{X_{\rm th}}{2}\right)e^{-\frac{X_{\rm th}}{2}}=P_{\rm FA}\,.
\end{equation}

where the threshold $X_{\rm th}$ is determined implicitly from $P_{\rm FA}$.
If $X$ now has a nonzero $\rho^2$, its probability of overcoming the threshold
becomes the detection probability, or 
\begin{equation}
\label{Pde}
P_{\rm DE} = P[X>X_{\rm th}|\rho^2]\,.
\end{equation}
The threshold $\rho_*^2$ is determined by requiring that when
$\rho^2\ge\rho_*^2$, Eq.~\eqref{Pde} provides a significant $P_{\rm DE}$.

\section{The detection statistic of multiple pulsars}
\label{f_statistic}
In this section we  extend the single-pulsar analysis approach of
Sec.~\ref{subsecthreshold} to apply to the detection of GWs from an ensemble of
pulsars.

\subsection{General theory}

To formulate how we might detect a combination of $n$ nearby sources, let us
consider the general problem of distinguishing the distribution of $n$ random
variables, $(X_1,\ldots,X_n)$ , between two probability densities $p_A$ and
$p_B$. Suppose we have a region $\mathcal{V}$, and we claim $A$ if
$(X_1,\ldots,X_n)\in \mathcal{V}$, and $B$ otherwise. In the context of GW
detection $A$ is without signal, while $B$ is detection. In this way, our
false-alarm probability is 
\begin{equation}
P_{\rm FA} = \int_{\bar{\mathcal{V}}} p_A(x_1,\ldots,x_n) dx_1\ldots dx_n,
\end{equation}
where $\bar{\mathcal{V}}$ represents not being within the region $\mathcal{V}$,
and our detection probability is
\begin{equation}
P_{\rm DE} = \int_{\bar{\mathcal{V}}} p_B(x_1,\ldots,x_n) dx_1\ldots dx_n.
\end{equation}

We then have to find the region $\mathcal{V}$ for which $P_{\rm DE}$ is
maximized given $P_{\rm FA}$. It is possible to find that the boundary of
$\mathcal{V}$ should be given by
\begin{equation}
\label{eqbound1}
\mu p_A (x_1,\ldots,x_n)= p_B (x_1,\ldots,x_n)\,.
\end{equation}
This is an implicit formula: given different values of the Lagrange multiplier
$\mu$, we  arrive at regions that have particular pairs of ($P_{\rm
FA},P_{\rm DE}$).  For each pair, the detection probability is the maximum
possible value given $P_{\rm FA}$.  Operationally, the boundaries of all these
$\mathcal{V}$'s are given by surfaces specified by Eq.~\eqref{eqbound1}. In
other words, for data $X_{1,\ldots,n}$, if we define the likelihood ratio
\begin{equation}
\mathcal L =\frac{p_B(X_1,\ldots,X_n)}{p_A(X_1,\ldots,X_n)}
\end{equation}
as a detection statistic, and by imposing a threshold, we  obtain the best
$P_{\rm DE}$ with given $P_{\rm FA}$.

If we have various versions of $B$ parameterized by a set of parameters
$\boldsymbol{\theta}$, we can further average over these possibilities with
their prior probability distributions  $w(\bg{\theta})$, such that 
\begin{equation}
P_{\rm DE} = \int d\bg{\theta}  w(\bg{\theta} ) \int_{\bar{\mathcal{V}}} p_B(x_1,\ldots,x_n;\bg{\theta} ) dx_1\ldots dx_n.
\end{equation}
This simply arrives at modified boundaries of $\mathcal{V}$ given by 
\begin{equation}
\label{eqbound}
\mu p_A (x_1,\ldots,x_n)= \int w(\bg{\theta} ) p_B (x_1,\ldots,x_n;\bg{\theta} ) d\bg{\theta} \,
\end{equation}
meaning that
\begin{equation}
\mathcal L =\frac{\displaystyle\int w(\bg{\theta} )p_B(X_1,\ldots,X_n;\bg{\theta} )d\bg{\theta} }{p_A(X_1,\ldots,X_n)}.
\end{equation}
This is in fact the same as the  marginal likelihood ratio ( the Bayes
factor in a Bayesian approach) for obtaining the data $X_{1,\ldots,n}$ ---
therefore we have simply established the optimality of the Neyman-Pearson
approach in our case. 



\subsection{Multiple pulsars}

In the detection of multiple pulsars, let us consider $A$ to be $n$ independent
4-D $\chi^2$ distributions, and $B$ to be $n$ independent 4-D noncentral
$\chi^2$ distributions, with noncentrality parameter $\lambda_1$, \ldots,
$\lambda_n$ (for simplicity, we use $\lambda$ rather than the optimal SNR
$\rho^2$).  Recall that for a $k$-D noncentral $\chi^2$ distribution, we have
\begin{equation}
p_{(k,\lambda)}(x) = \frac{1}{2}e^{-(x+\lambda)/2}
\left(\frac{x}{\lambda}\right)^{k/4-1/2} I_{k/2-1}(\sqrt{\lambda x})\,,\quad x>0\,.
\end{equation}
where $I$ is the modified Bessel function of the first kind. We can then write
\begin{equation}
p_A (x_1,\ldots,x_n) = p_{(4,0)}(x_1) \cdots p_{(4,n)}(x_n)
\end{equation}
and
\begin{equation}
p_B (x_1,\ldots,x_n) = p_{(4,\lambda_1)}(x_1) \cdots p_{(4,\lambda_n)}(x_n)
\end{equation}

Following Eq.~\ref{eqbound1}, for fixed values of $\lambda_1$, \ldots
$\lambda_n$, we have
\begin{equation}
 \prod_{j=1}^n \frac{ 2e^{-\lambda_j/2} I_1\left(\sqrt{\lambda_j x_j}\right)}{\sqrt{\lambda_j x_j}} =\mu
\end{equation}
as optimal boundaries of $\mathcal{V}$ which can also be written as
\begin{equation}
\sum_j \log \left[\frac{I_1\left(\sqrt{\lambda_j x_j}\right)}{\sqrt{\lambda_j x_j}}\right] = \mbox{const}\,.
\end{equation}
This shows how signals should be combined resulting in our combined detection
statistic
\begin{equation}
\label{optfixed}
\mathcal L^{\rm opt}_{\rm fix}= \sum_j \log
\left[\frac{I_1\left(\sqrt{\lambda_j X_j}\right)}{\sqrt{\lambda_j X_j}}\right]
\end{equation}
where $X_{1,\ldots,n}$ are the $n$ observables.

If each $\lambda$ depends on a set parameters $\boldsymbol{\theta}$, and for
each $j$ there is a corresponding prior distribution $w_j(\bg{\theta})$, then
from Eq.~\eqref{eqbound}, we can write
\begin{equation}
\label{eqoptglobal}
\mathcal L^{\rm opt}= \sum_j \log\left[ \int w_j(\bg{\theta}) e^{-\lambda(\bg{\theta})/2}  \frac{I_1\left(\sqrt{\lambda(\bg{\theta}) X_j}\right) }{\sqrt{\lambda(\bg{\theta}) X_j}}d\bg{\theta}\right].
\end{equation}
As a sanity check, if $w_j(\lambda)=\delta(\lambda-\bar\lambda_j)$, we recover
the previous result. 

%
%

\subsection{Special case: exponential distribution}
\label{subsec:exp}

We can further simplify the construction of the optimal statistic,  simply
and arbitrarily assuming that each $\lambda_j$ value is drawn from an
exponential distribution, or 
\begin{equation}
w_j  (\lambda) = \frac{1}{\bar\lambda_j}e^{-\lambda/\bar\lambda_j} \,,\quad \lambda>0\, ,
\end{equation}
where $\bar{\lambda}$ is the mean value of the prior distribution on $\lambda$
for each pulsar. In this case, we obtain the following closed-form
expression,
\begin{equation}
\label{lexpopt}
\mathcal{L}_{\rm exp}^{\rm opt} =  \sum_j \log\left[\frac{e^{Y_j}-1}{Y_j}\right]
\end{equation}
with
\begin{equation}
Y_j = \frac{\bar\lambda_j}{\bar\lambda_j+2}\frac{X_j}{2}
\end{equation}
This is quite interesting: those sources with $\bar\lambda_j \gg 2$ (already
quite detectable individually), should be combined with a similar weight, while
those much less than unity should be combined according to the expectation
value of the noncentrality parameter, $\bar{\lambda}_j$.  The latter case is 
discussed further below.  

\subsection{Special case: weak-signal limit}

A different way to obtain an optimal statistic is to directly assume that we
should linearly combine the $\mathcal{F}$-statistic according to
\begin{equation}\label{lin_opt}
\mathcal{L}^{\rm wopt}_{\rm lin} =\sum_j \alpha_j X_j\,,
\end{equation} 
and optimize the ``signal-to-noise ratio'', which is given by the increase of
$\langle\mathcal{L}^{\rm wopt}_{\rm lin}\rangle$ due to nonzero $\lambda_j$
divided by the variance of $\mathcal{L}^{\rm wopt}_{\rm lin}$ in the absence of
signal.  This leads to 
\begin{align}
\alpha_j & \propto \bar{\lambda}_j \nonumber  \\
&=\frac{f_j^4 \langle \mathcal{K}_j \rangle}{d_j^2 S_{h}(2f_j)},
\end{align}
where the second line  is  valid for the known pulsars case, if we assume the  intrinsic parameter of  pulsars follows the same distribution (see the  discussion in Sec. \ref{sim_known} ). 
This can be derived from the optimal statistic, if we assume that we are
interested in the low signal amplitude limit where the $\lambda_j$ are small.
In this case, we can Taylor expand Eq.~\eqref{eqoptglobal} and obtain, at
leading order
\begin{equation}
\mathcal{L}^{\rm wopt}_{\rm lin} \approx \sum_j \frac{(X_j-2)\bar{\lambda}_j}{4}
\end{equation}
which is equivalent to using 
\begin{equation}
\mathcal{L}^{\rm wopt}_{\rm lin} \approx \sum_j \alpha_j X_j , 
\end{equation}
which is also consistent with Eq.~\eqref{lexpopt} when $\bar\lambda_j$ is small. 

This implies that if we could tolerate a high false-alarm probability by
setting our threshold low, it is plausible that combining the observables
proportional to the (prior) expectation value of noncentrality parameters
would be optimal.  However, as shown in Sec.~\ref{sim}, in the situations we
encounter, this approximation is not quite valid. 

\subsection{Comparison with individual pulsar detection}
Before we compare our strategy with existing strategies that do not combine
signals from multiple pulsars, let us first clarify what it means to ``not
combine signals''.  A careful examination provides two possible variants.  

\subsubsection{Assigning equal false-alarm probability to each pulsar}
\label{subequalth}

The first approach regards treating each pulsar as truly independent, and by
setting the same false-alarm probability for each pulsar --- even though each
pulsar is not equally likely to provide detection.  In this procedure, we
therefore set the same threshold $X^{\rm th}$ for each pulsar, requiring
\begin{equation}\label{fa_individual}
1- P^n(X<X^{\rm th}|\rho^2=0)= P_{\rm FA}
\end{equation}
and leading to the following total detection probability 
\begin{equation}
P_{\rm DE}=1 -\prod_{j=1}^n P(X<X^{\rm th}|\lambda_j)
\end{equation}
of detecting at least one pulsar within this ensemble.

\subsubsection{Assigning false-alarm probability according to signal strength}
\label{subvaryth}
This is clearly problematic since we have potentially $\sim 200$ pulsars ---
assigning the same false-alarm value to pulsars with dramatically different
potential signal strength is clearly wasteful. If a different threshold is set
for each pulsar, in such a way that the detection probability of an ensemble is
maximum, we  then require
\begin{align}
\label{muthreshold}
\mu = &\int w_j(\bg{\theta})\frac{p_{(4,\lambda_j)}}{p_{(4,0)}} d\bg{\theta}  \nonumber\\
=&\int w_j(\bg{\theta})\frac{e^{-\lambda(\bg{\theta})/2} I_1\left(\sqrt{\lambda(\bg{\theta}) X_j^{\rm th}}\right)}{\sqrt{\lambda(\bg{\theta}) X_j^{\rm th}}} d\bg{\theta} 
\end{align}
where $\mu$ is a constant independent of $j$. As we vary $\mu$, we obtain a
varying set of $X_j^{\rm th}$ that would provide us with the optimal thresholds
for each $X_j$, such that the total detection probability of detecting a GW
signal within this ensemble is maximum given the false-alarm probability [as
defined in Eq.~\ref{fa_individual} with different $X^{\rm th}$ ].

\begin{figure}
\centerline{\includegraphics[width=0.5\textwidth]{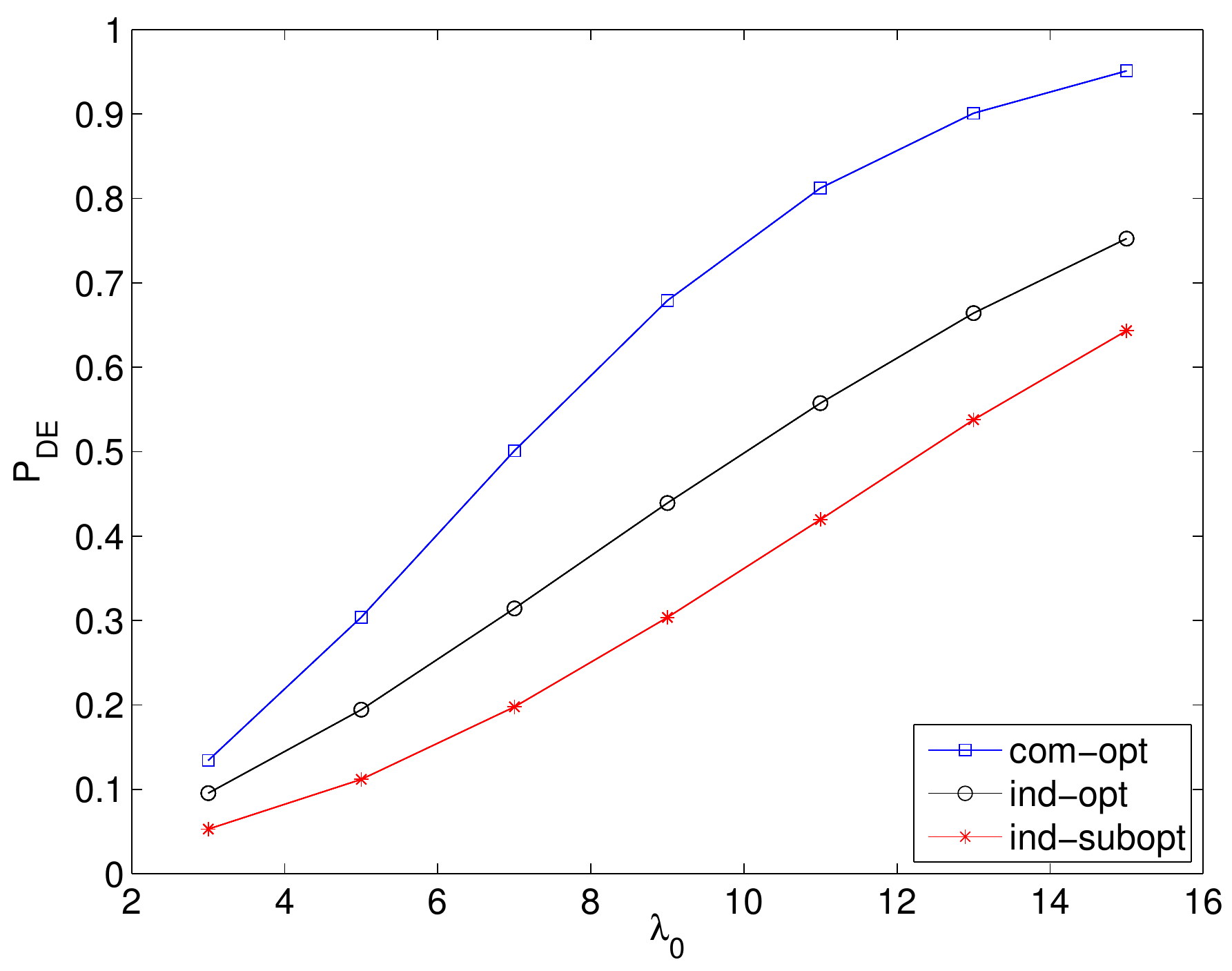}}
\caption{(Color Online.) Detection probability for the model given by
Eq.~\eqref{lambdasample}, using the optimal combined statistic (com-opt) and
individual thresholds (ind-subopt for equal thresholds, see
Sec.~\ref{subequalth}, and ind-opt for optimal thresholding, see
Sec.~\ref{subvaryth}).  We have fixed $P_{\rm
FA}=0.01$.\label{fig:fixedlambda}}
\end{figure}


\section{Monte Carlo Simulations of Simple Models}
\label{sim}
In this section, we perform numerical investigations of two simple models. In
particular, we  study the case of constant  $\lambda_j$ first, then the
case where the $\lambda_j$ values follow exponential distributions. This provides important basic understanding before we move on to the known pulsars.

\subsection{Constant $\lambda_j$ }

In this section, we  perform Monte Carlo simulations for signals with
fixed $\lambda_j$ -- the simplest case.  We shall compare four strategies: (i)
imposing a constant threshold on all $X_j$ [Sec.~\ref{subequalth}], (ii)
imposing a variable threshold on $X_j$, according to Eq.~\eqref{muthreshold}, (iii) using a linear-combination statistic
\begin{equation}
\label{eqlincom}
\mathcal{L}_{\rm lin} = \sum_{j=1}^n\alpha_j^\beta X_j
\end{equation}
with various values of $\beta$, and (iv) using  the optimal statistic, according to Eq.~\eqref{optfixed}. 
\begin{figure}
\centerline{\includegraphics[width=0.5\textwidth]{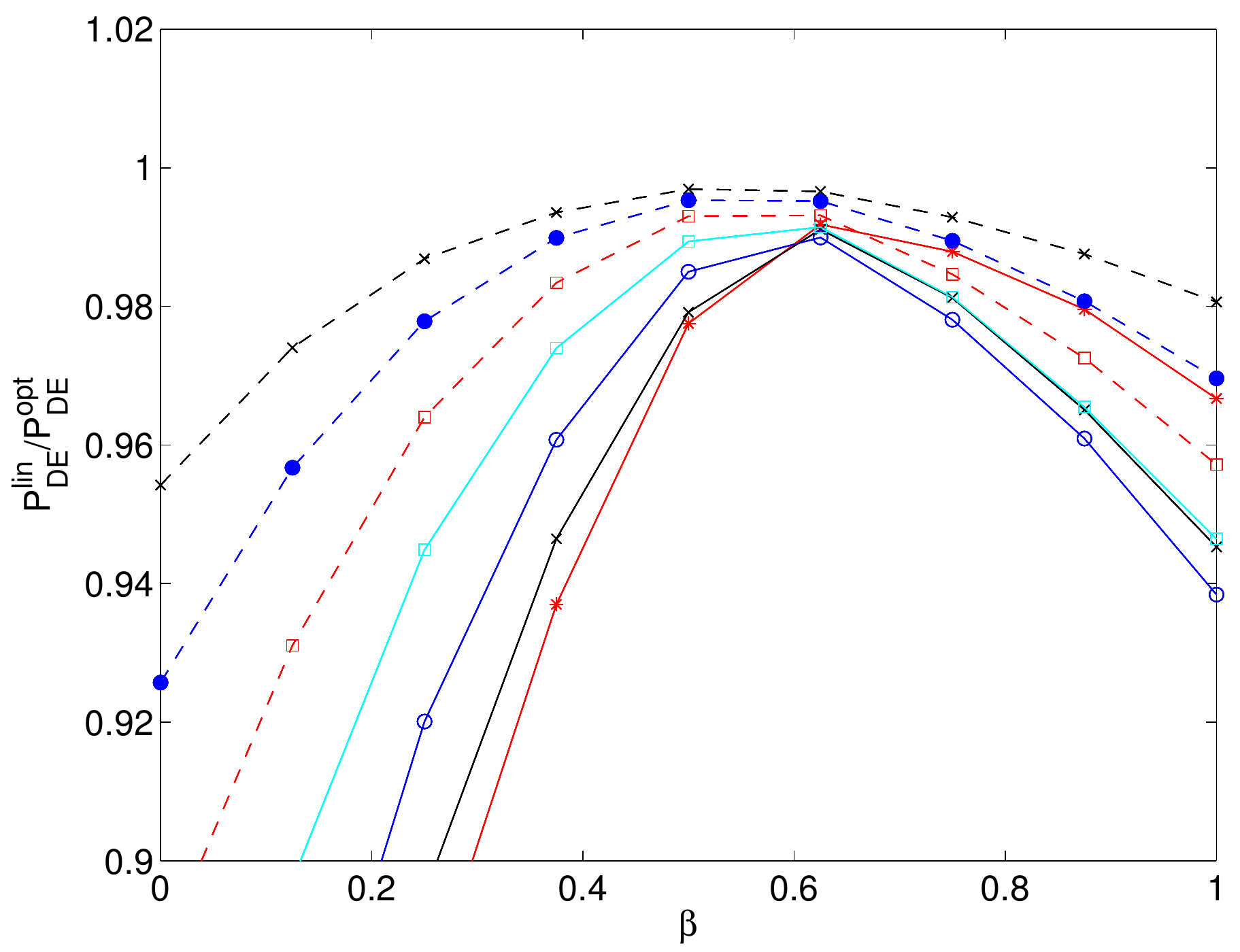}}
\caption{(Color Online.) Detection probability of linear combination
statistics [Eq.~\eqref{eqlincom}] compared to the optimal detection
probability. Different traces correspond to $\lambda_0$ ranging from 3 to 10,
and we have fixed $P_{\rm FA}=0.01$. \label{fig:beta}}
\end{figure}

We have chosen
\begin{equation}
\label{lambdasample}
\alpha_j=\lambda_j = \frac{\lambda_0}{j},\quad j=1,\ldots,8,
\end{equation}
which is designed to simulate an ensemble of sources that are distributed
on a two-dimensional plane. If, within each disk with radius $r$, the number of sources
is proportional to $r^2$, then for the $N$-th source, its distance should
be ${\sim}\sqrt{N}$; therefore the noncentrality parameter should be ${\sim}1/N$.

As we vary $\lambda_0$ from 3 to 15, and fixing $P_{\rm FA}=0.01$, we compare
the detection probability. As is shown by Fig.~\ref{fig:fixedlambda}, the
optimal strategy is substantially better than strategies (i) and (ii).  In
particular, in order for (i) and (ii) to achieve 50\% detection probability,
the noncentrality parameter must be a factor of $\sim 2$ stronger.  


In Fig.~\ref{fig:beta}, we investigate the performance of the
linear-combination statistics.  For $P_{\rm FA}=0.01$, we plot $P_{\rm DE}$ as
a function of the index $\beta$.  It seems here that $\beta\sim 0.5$ performs
slightly better than $\beta \sim 1$, although the optimal $\beta$ value depends
on $\lambda_0$, and is located somewhere between 0.5 and 0.8.

\subsection{Exponential distributions for $\lambda_j$}

Let us now consider $\lambda_j$ values that have simple exponential prior
distributions for which we have analytical formulas derived in Sec.
~\ref{subsec:exp}.  This is  also important because we can test whether
having the correct prior information in constructing the detection statistic
can significantly affect detection efficiency.  In particular, while the
optimal statistic seems highly dependent on the prior distribution of
$\lambda$, the linear statistic $\mathcal{L}_{\rm lin}$ is robust against a
rescaling of the distributions of all $\lambda_j$. 

\begin{figure}
\includegraphics[width=0.5\textwidth]{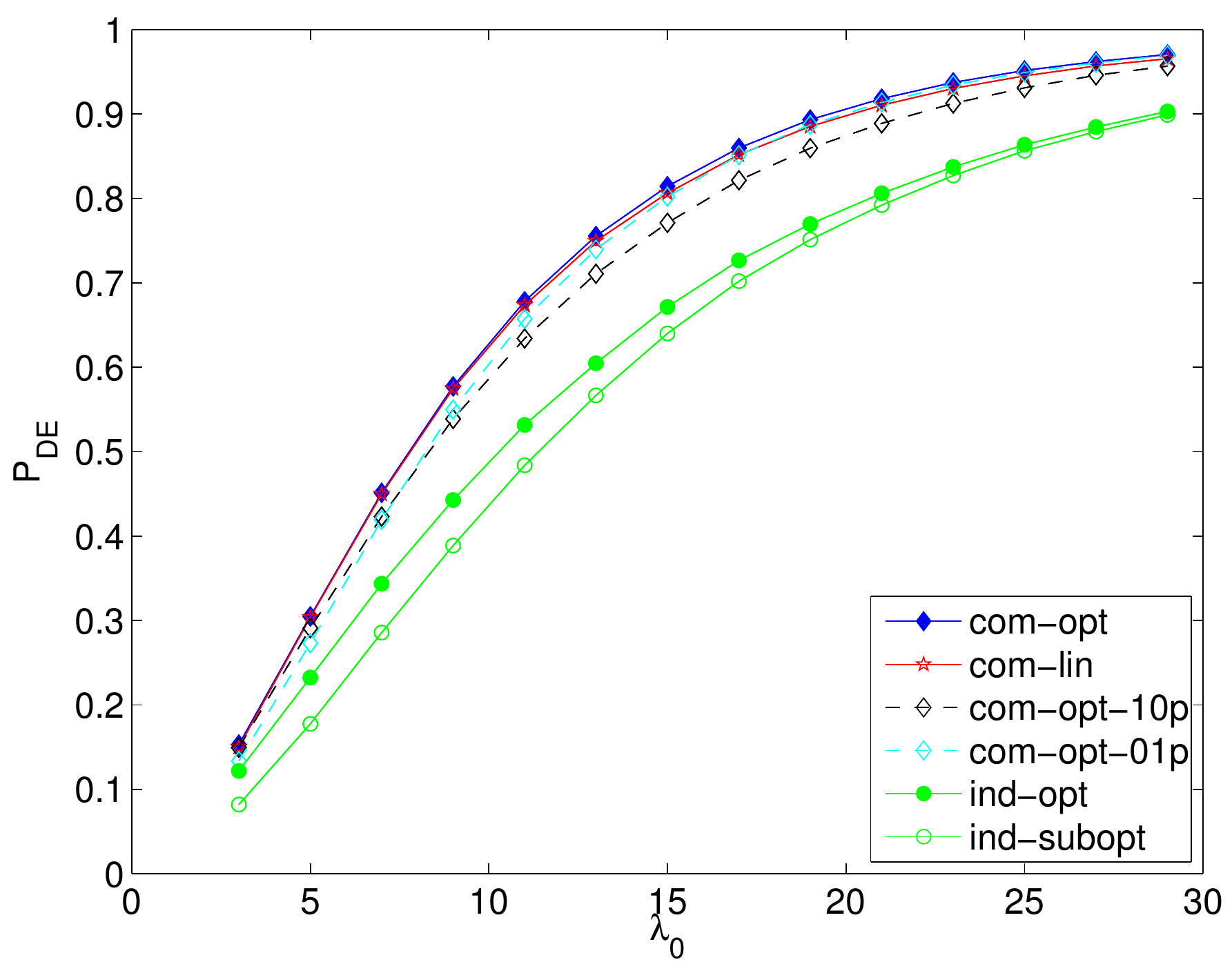}
\caption{Detection probability for models with $\lambda_j$
following exponential distributions with mean value given by
Eq.~\ref{lambdasample}. We have fixed $P_{\rm FA}=0.01$. Shown here are from
the optimal statistic (com-opt), optimal statistic scaling prior distributions
by 10 (com-opt-10p) and by 1/10 (com-opt-01p), linear-combination statistic
with $\beta=1/2$ (com-lin), individual pulsar detection with optimal
thresholding on each $X_j$ (ind-opt) and individual pulsar detection with
common threshold (ind-subopt). 
\label{fig8exp}}
\end{figure}

Again, to be concrete, we chose  to have $\lambda_j$'s follow exponential
distributions, with mean values given by Eq.~\ref{lambdasample}. The detection
probability with $P_{\rm FA}=0.01$ is shown in Fig.~\ref{fig8exp} for
$\lambda_0$ ranging from 3 to 30. Here, we see again that the optimal statistic
is substantially better than individually detecting the pulsars --- while a
more strategic thresholding allows some improvement. 

In this case, we can see the potential benefits of the linear statistic: when
the wrong prior distributions are used (with $\bar\lambda_j \rightarrow
10\bar\lambda_j$ and $\bar\lambda_j \rightarrow \bar\lambda_j/10$) to compute
the optimal statistic, the detection efficiency drops to a level worse than
using the linear statistic, which is independent of an overall rescaling of
all $\lambda_j$ values. 

\begin{figure}
\includegraphics[width=0.5\textwidth]{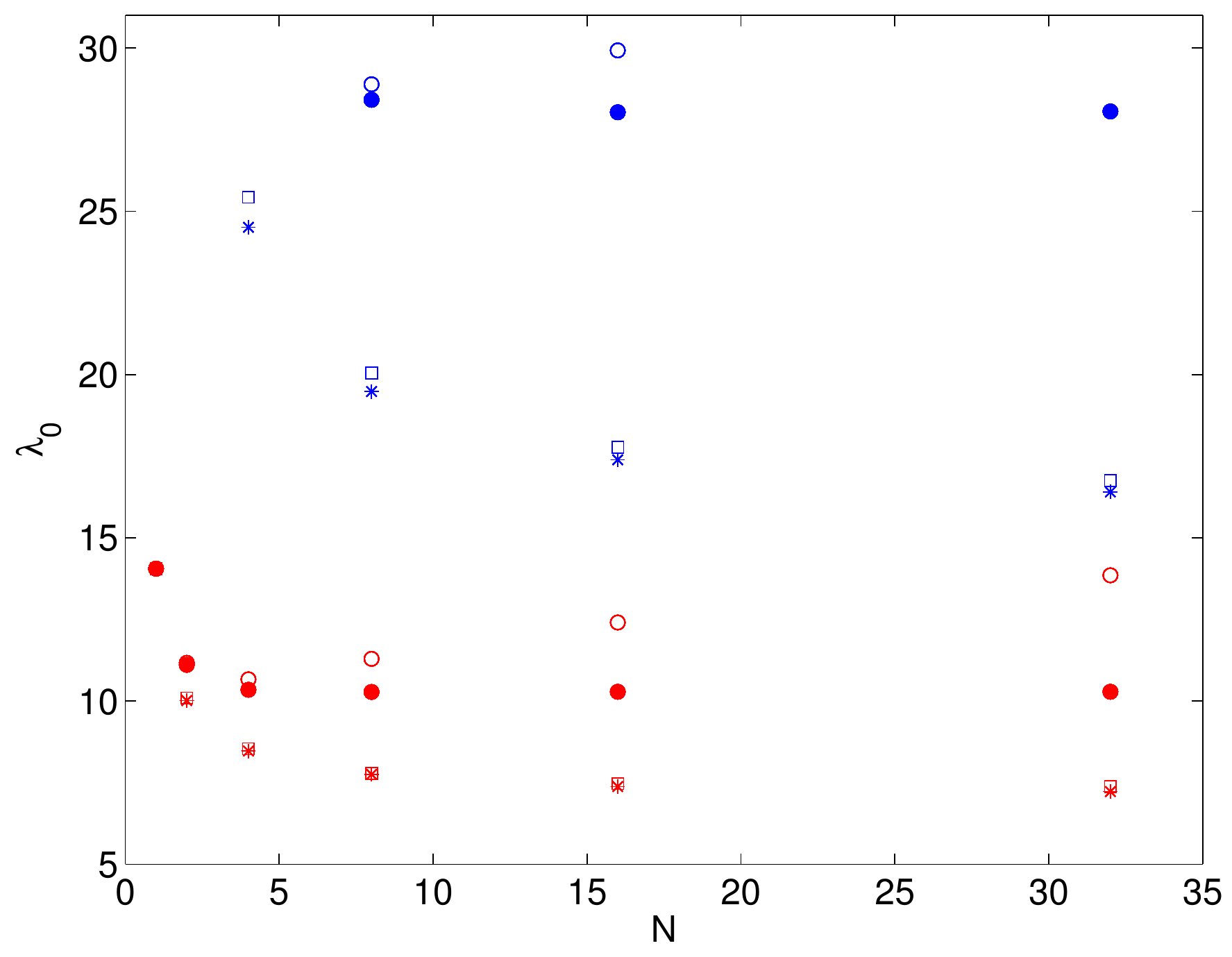}
\caption{Values of signal strength ($\lambda_0$) at which each
detection strategy can yield detection probability of 50\% (red symbols) and
90\% (blue symbols), respectively, as a function of the number of sources, $N$.
We have used the optimal statistic (stars), linear-combination statistic with
$\beta=0.5$ (square), uniform threshold for all sources (hollow circles) and
optimal thresholding (solid circles).  Increase of $\lambda$ with $N$ in the
uniform threshold case indicates that including more sources introduces
contamination from weaker sources. We assumed exponential distributions for
$\lambda_j$ in this plot and have fixed $P_{\rm FA}=0.01$. 
\label{fig:scaling}}
\end{figure}

\subsection{Scaling with the number of sources}
Let us now consider how the detection probabilities of the various schemes
scale with the number of sources. We do this by simply extending
Eq.~\eqref{lambdasample} to include a variable number of sources N.
 
In Fig.~\ref{fig:scaling}, we can see that as the number of sources increases
the detection probability of the optimal and linear combination statistics with
$\beta=0.5$ also increases. The detection probability of individual
pulsars using a common threshold decreases, while the individual detection with
optimal thresholding also keeps increasing, but stops increasing at a
relatively low number of sources. This can be explained as being due to the combined
statistics' ability to incorporate weaker sources without sacrificing
sensitivity.

Numerically, we can see that a substantially larger signal strength has to be
present for the individual detection strategies. In addition, we emphasize
that the linear-combination statistic, here shown to be very close to being
optimal, is independent from an overall rescaling of the distribution of
$\lambda_j$'s. The optimal thresholding, on the other hand, does depend on the
particular model of $\lambda$. 

\section{Monte Carlo Simulations for Known Pulsars}\label{sim_known}
 
We  now discuss the case of detecting GWs from multiple known pulsars. We
start by describing the known and unknown aspects of these sources, and
then present the setup and conclusions of our numerical simulations. 

\subsection{Known pulsars: prior distributions for $\lambda_j$.}
For the case of multiple pulsars, the noncentrality parameter $\lambda_j$ for
each pulsar in a single detector is simply equal to $\rho_j^2$, as given by
Eq.~\eqref{eqrho}. We  now discuss in detail all factors contributing to
our prior knowledge of $\rho_j^2$.  

The ellipticity $\epsilon_j$ crucially defines the level of quadrupole
deformation of the NS. At present, we have only theoretical constraints based
on the internal structure of NSs, which span a wide range, and observational
upper limits from from previous GW searches, which span the range $\sim
10^{-7}-10^{-2}$~\citep{2010ApJ...713..671A}. Our baseline assumption is that
the $\epsilon$ of all pulsars follows a common (yet unknown) distribution; this
could be motivated as arising from the belief that all these eccentricities
were generated by the same physical mechanism. We note that it is plausible for
Advanced LIGO to detect at the level of $\overline{\epsilon} \sim
\mbox{few}\times 10^{-8}$.  

%
%


The geometrical factor $\mathcal{K}$ depends on the inclination angle $\iota$,
polarization angle $\tilde\psi$, and  antenna patterns $F_{++}$ and
$F_{\times\times}$, see Sec.~\ref{sec:fstat}. We  assume no knowledge
concerning the orientation of the pulsar, therefore uniformly distributing
$\cos\iota$ between $-1$ and $+1$, and uniformly distributing $\tilde\psi$
between $0$ and $2\pi$.  As for $F_{++}$ and $F_{\times\times}$, they further
depend on the geographical location and orientation of the detector, as well as
the source's declination angle (the right ascension dependence is averaged away
after many sidereal days observation).  

  

As noted by CS, for the network of $M$ detectors case, the noncentrality
parameter of each pulsar is simply
\begin{equation}
\lambda_{j}^{\rm net} =\sum_i^M \lambda_{ji},
\end{equation}


\subsection{Simulations and results}
Our simulations assume one year of observation using the network of Advanced
LIGO and Virgo at design sensitivity~\footnote{LIGO Document
T1200307-v4}~\footnote{LIGO Document T1300121-v1}. The positions and
orientations of the detectors are taken from Table 1
of~\cite{1998PhRvD..58f3001J}.  The known pulsar parameters (distance, sky
location and frequency) used to compute $\lambda_{j}^{\rm net}$ are taken from
the 195 known pulsars analysed in the initial detection era, and we assume that
the moment of inertia $I=10^{38} \, \rm{kg \,m^2}$~\citep{2014ApJ...785..119A}.
To provide a proof-of-principle of our proposed method, we assume (i)
$\epsilon^2$ values follow exponential distributions with two different rate
parameters $4\times10^{-16}$ and $9\times10^{-16}$, and (ii) $\epsilon$ values
follow normal distributions with mean values of $1.5\times10^{-8}$, and
$2\times10^{-8}$ with standard deviations equal to half of their respective
mean values.

We have performed simulations to test the detection efficiency of our proposed
robust statistic ${\mathcal L_{\rm lin}}$ (Eq.~\ref{eqlincom}, see discussion
in~\ref{sim}) via the receiver operating characteristic (ROC) curve, which is a
parametric plot of the probability of false alarm versus the probability of
detection.  

The ROC curve is constructed using $10^5$ simulations of $X_j$ with
noncentrality parameter $\lambda_{j}^{\rm net}$ and $10^5$ noise only
simulations. With the assumption of $\epsilon^2$ following the same
distribution,   the $\alpha_{j}^{\rm net}$ is  used in
Eq.~\ref{eqlincom} to compute every simulated $\mathcal L_{\rm lin}$ in place of $\alpha_{j}$ for the $M$ detectors case,  defined as 
\begin{equation}
\alpha_{j}^{\rm net} =\sum_i^M \alpha_{ji}=\sum_i^M \frac{f_j^4 \langle \mathcal{K}_j \rangle}{d_j^2 S_{hi}(2f_j)}.
\end{equation}
We compare the detection efficiencies of ensemble based
strategies including the weighted-combination ($\beta=0.5$) and
equal-combination ($\beta=0$, the CS case) method, with the individual
pulsar detection strategy including the expected brightest [the largest value
of $\frac{f^4 \langle \mathcal{K} \rangle}{d^2 S_h(2f)}$], measured brightest
(the maximum $\rho_j^2$ in each simulation) case.   

\begin{figure*}[t]
\begin{center}
\begin{tabular}{cc}
\includegraphics[angle=0,width=3in]{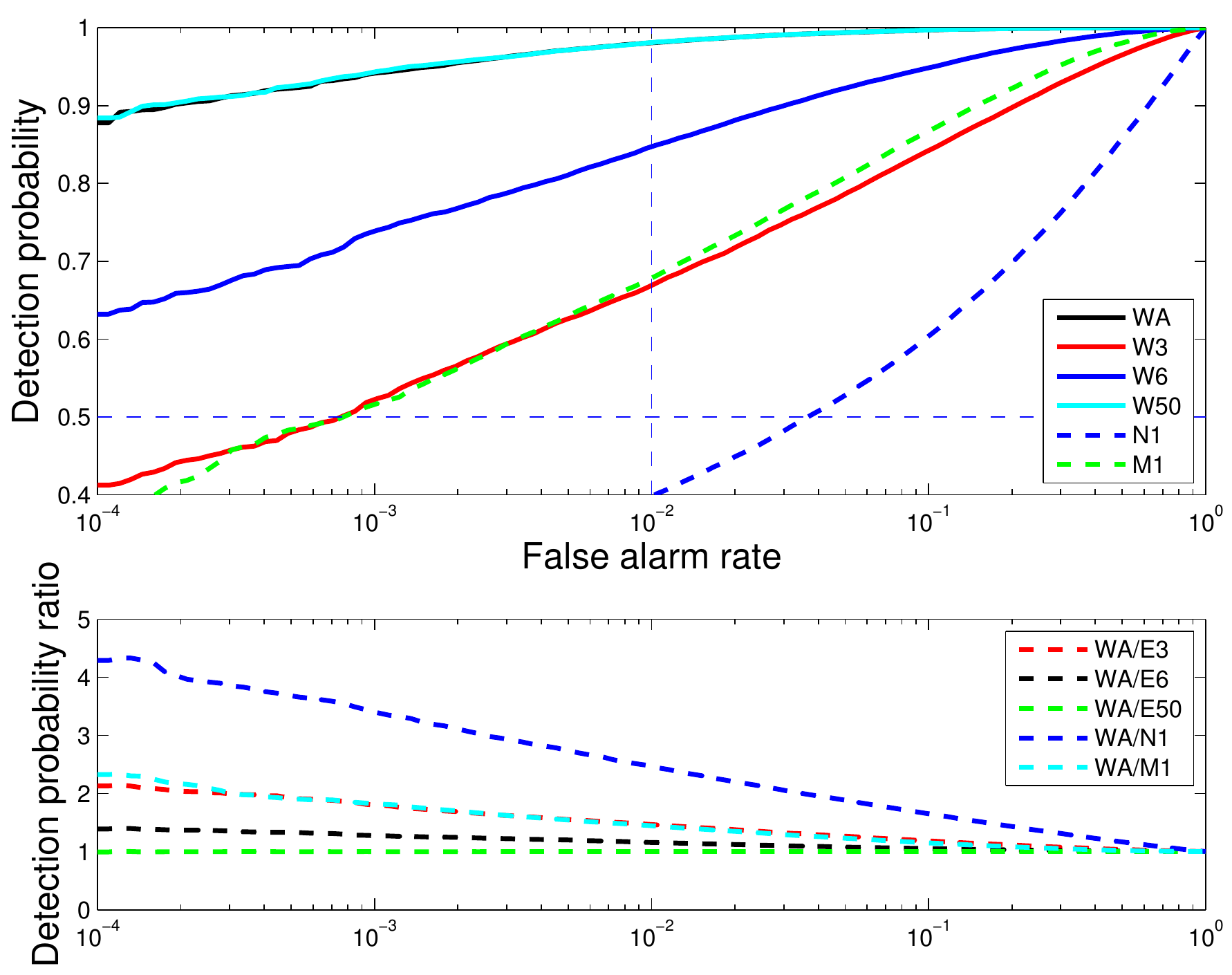}  
\includegraphics[angle=0,width=3in]{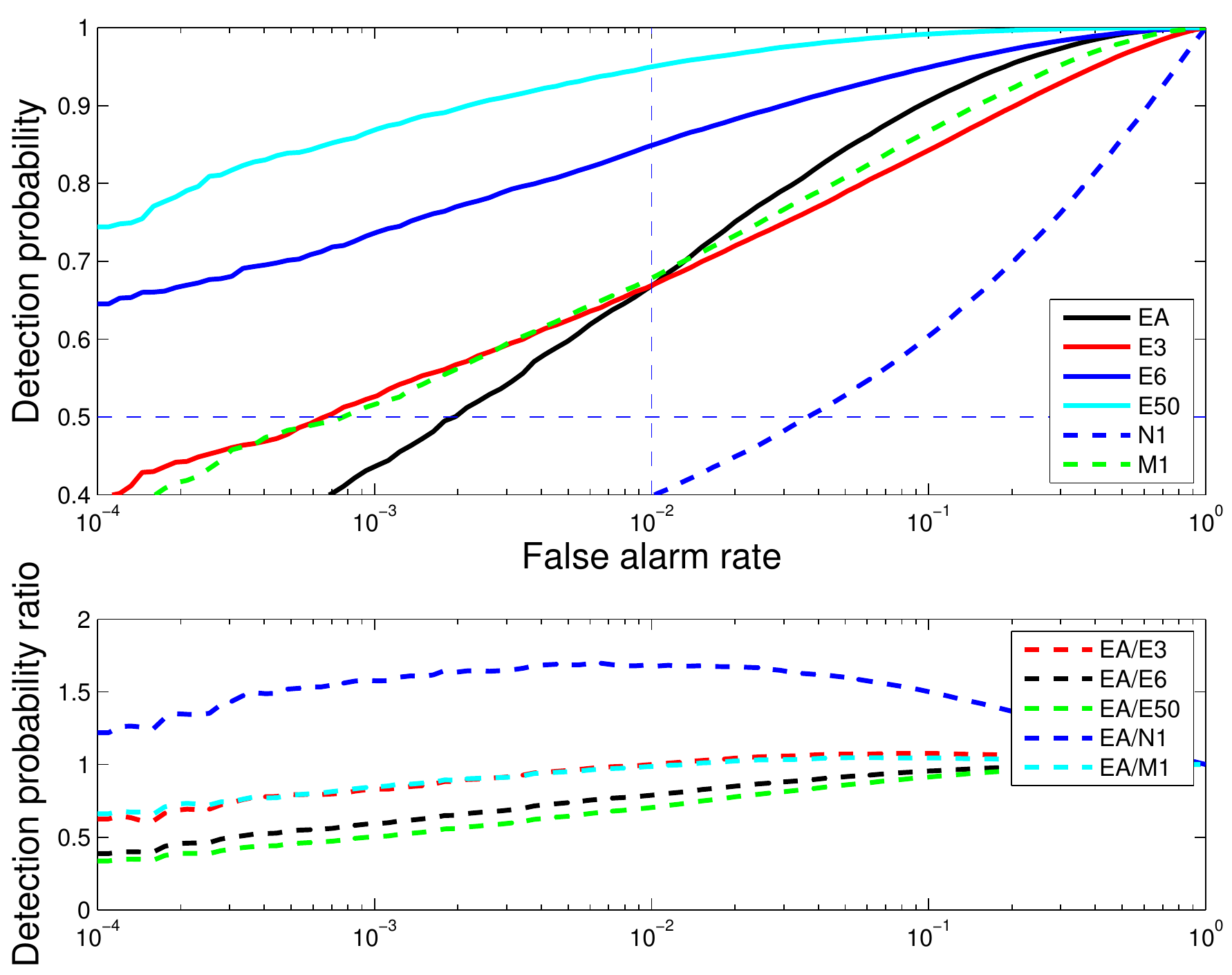} 
\end{tabular}
    \end{center}
\caption{ROC curves for  different detection methods. WA (EA), W3 (E3), W6
(E6), W50 (E50), N1 and M1 correspond to weighted (equal) combinations of all,
the expected brightest three, the expected brightest six,  the expected
brightest fifty, the expected brightest and the measured brightest
pulsar(s) respectively.  The  $\epsilon^2$ in injections  were drawn from}   an exponential
distribution with rate parameter $2\times10^{-16}$.
\label{roc_moresource_exp}.
\end{figure*} 
 \begin{figure*}[t].pdf \begin{center}
\begin{tabular}{cc}
\includegraphics[angle=0,width=3in]{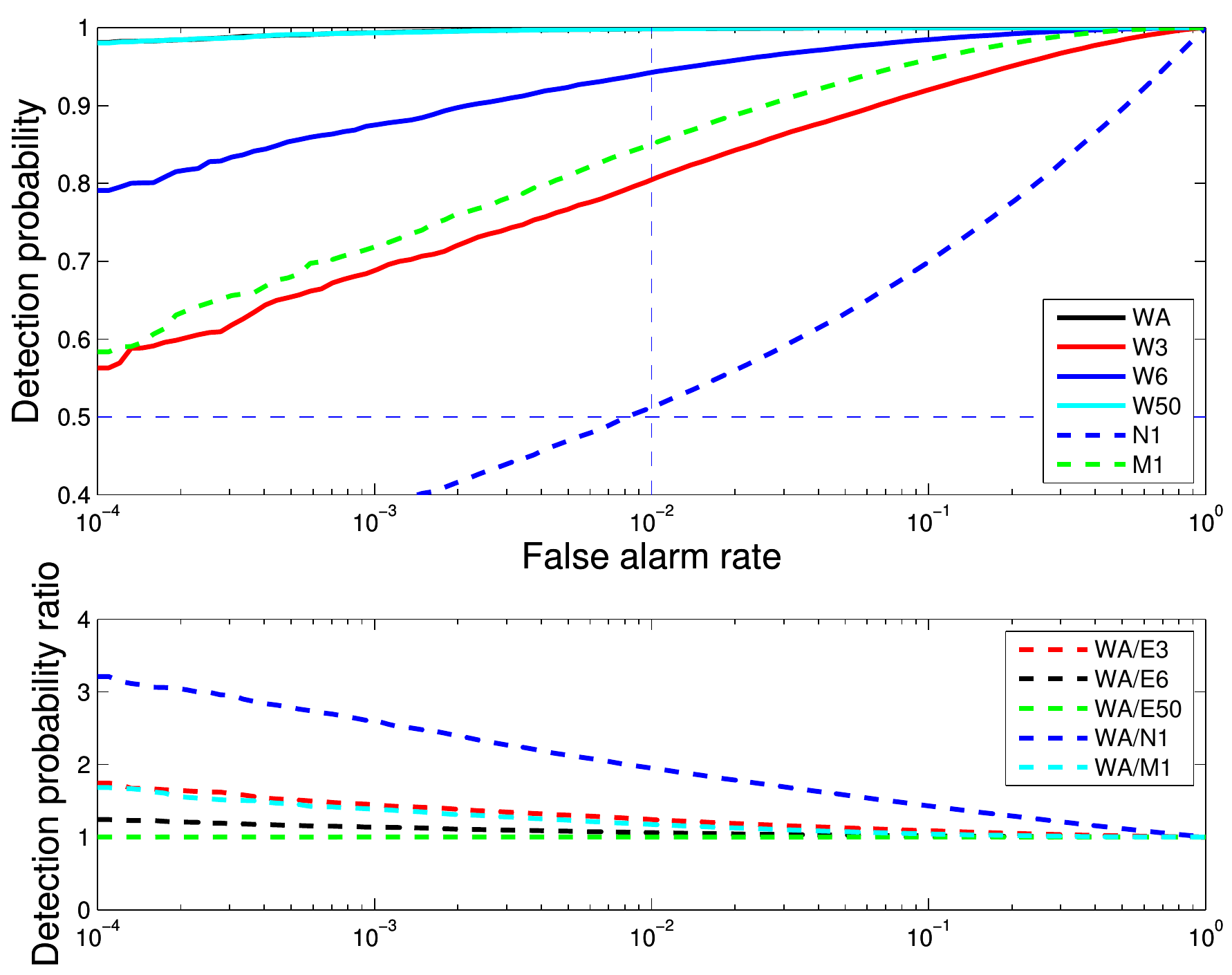}
\includegraphics[angle=0,width=3in]{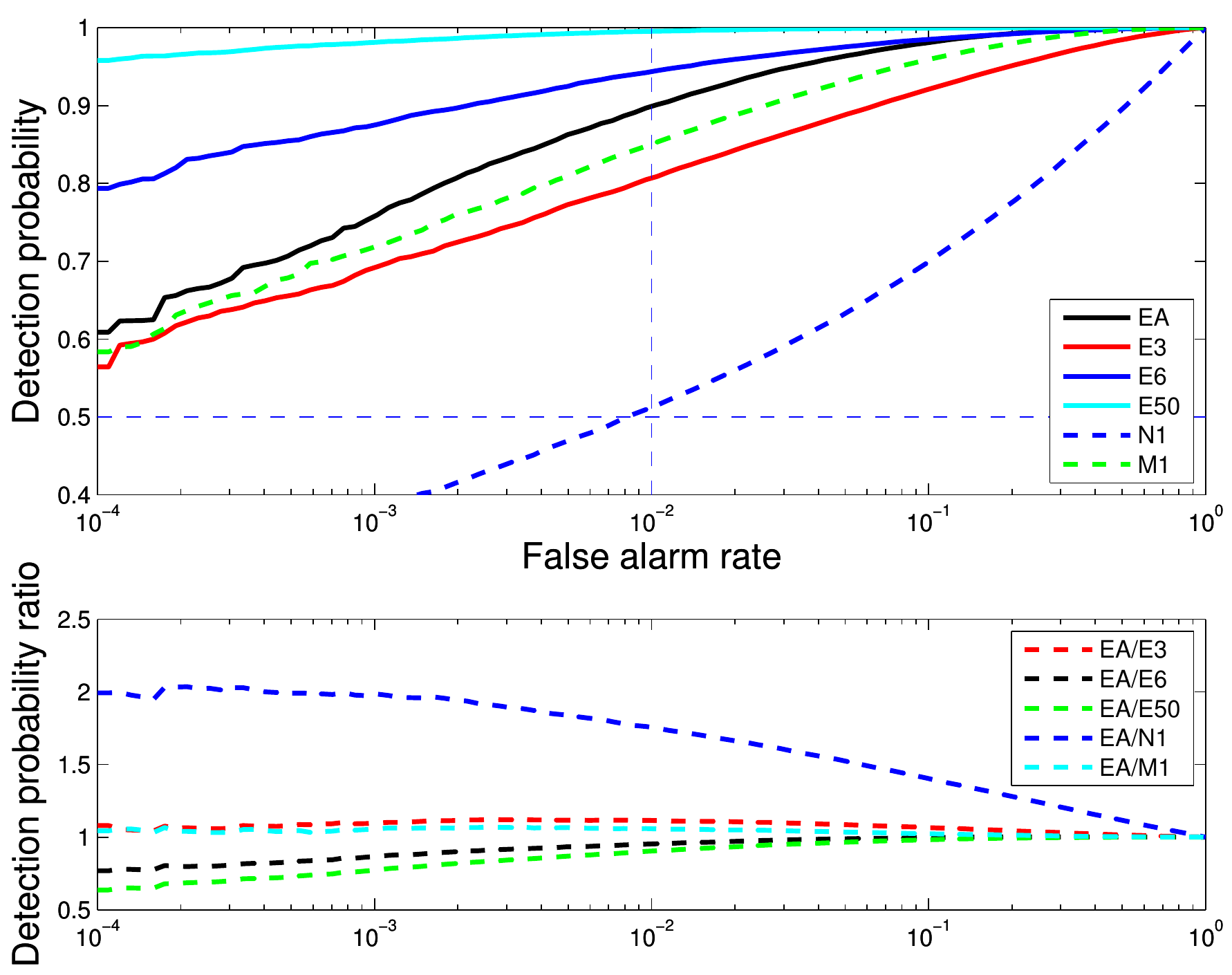}
\end{tabular}
 \end{center} 
 \caption{Same as Fig.~\ref{roc_moresource_exp}
but with  $\epsilon$ following the  Gaussian distribution with mean value of
$1.5\times10^{-8}$, and half  of  mean values as its variance value.
\label{roc_moresource_nor}}.  
\end{figure*} 
 
The results are first presented in terms of answers to the following two
questions: Will collecting more pulsars return higher $P_{\rm DE}$ values than
an individual detection? How many sources should be combined to obtain the
maximum $P_{\rm DE}$ at given $P_{\rm FA}$? As shown in
Figs.~\ref{roc_moresource_exp} and~\ref{roc_moresource_nor} for various
ellipticity distributions, the more sources are combined, the higher  $P_{\rm
DE}$ is for  our proposed robust statistic ${\mathcal L_{\rm lin}}$ with
$\beta=0.5$, although combining the weakest part of the population (e.g. the
weakest 50 sources) will not greatly contribute to $P_{\rm DE}$. As expected,
we find that the $P_{\rm DE}$ increases when combining the first few high
amplitude sources, and then decreases for the equal-weight method ($\beta=0$)
as more and more weak sources are added to the combination. These results are
consistent with the simple test in Sec.~\ref{sim}. Since we do not know the
true values of all pulsar parameters, it is interesting to ask whether the
measured brightest source or the expected brightest source would be more
detectable than any other ensemble of sources.

Since the weighted-combination method is optimized for the whole population of
GW signals, neither the measured brightest source nor the expected brightest
one is more detectable than the whole population. This is not the case for the
equal-weight combination method (see Figs.~\ref{roc_moresource_exp} and
\ref{roc_moresource_nor}). 

\begin{figure*}
 \begin{center}
\begin{tabular}{cc}
\includegraphics[angle=0,width=3in]{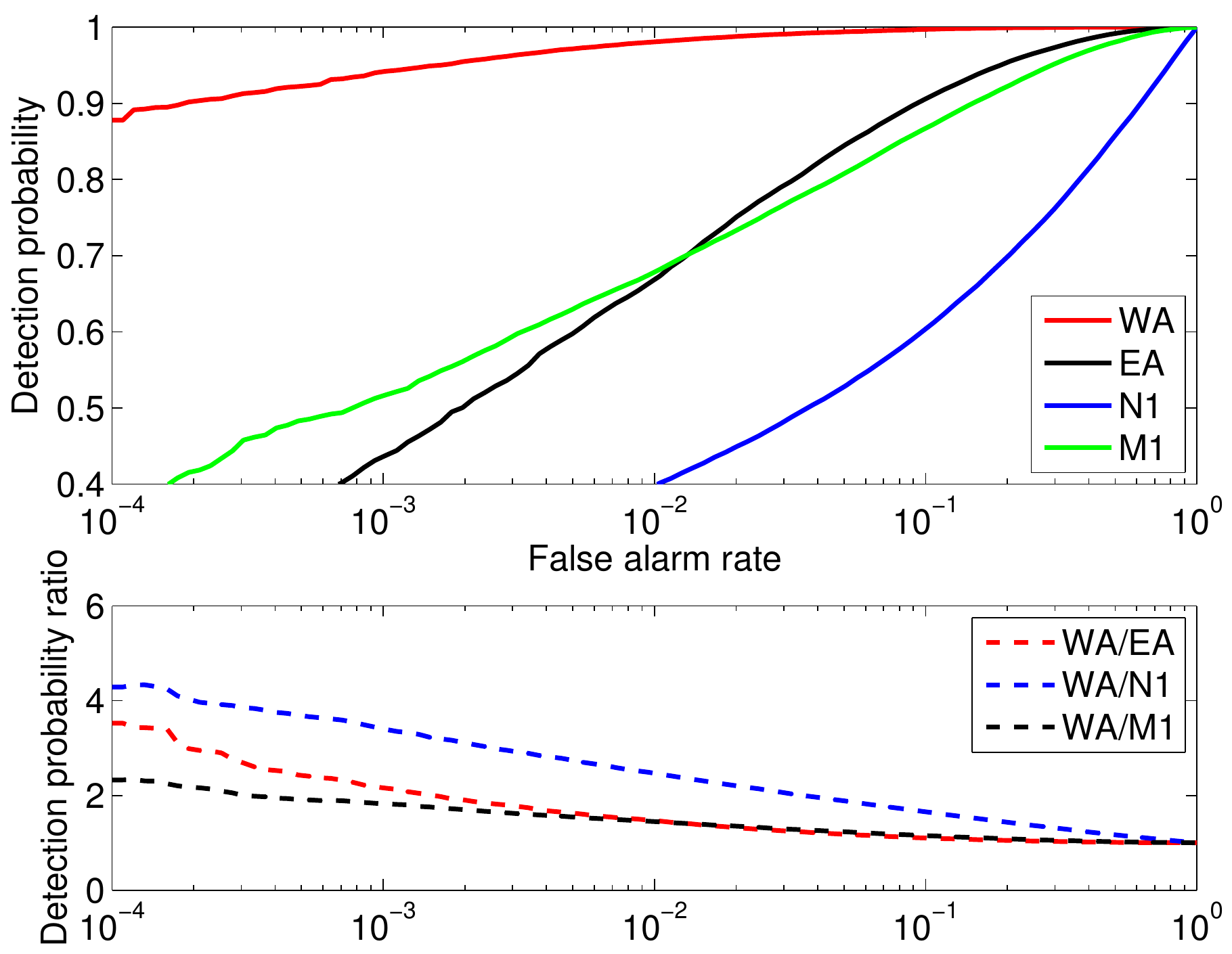}
\includegraphics[angle=0,width=3in]{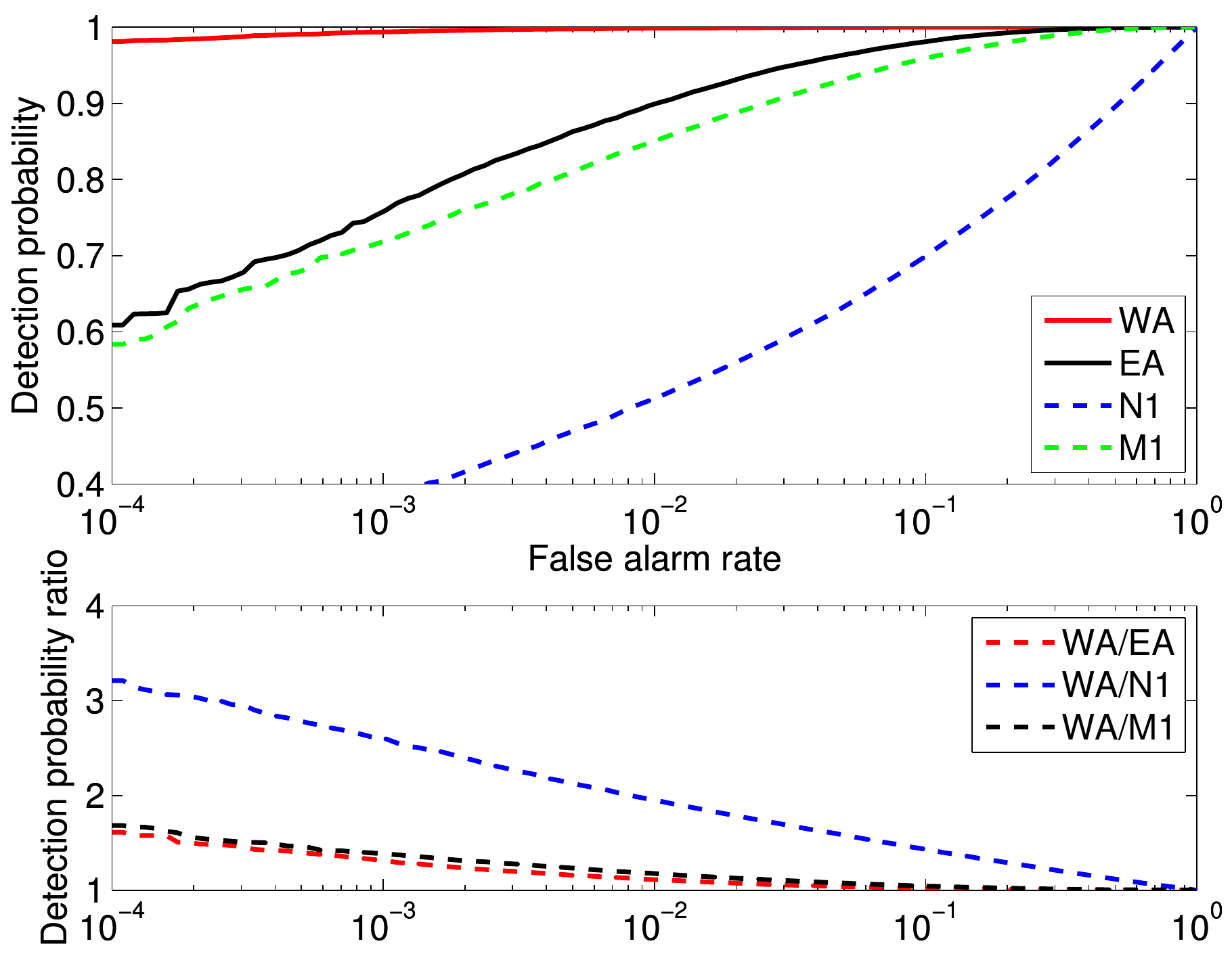}\\
\includegraphics[angle=0,width=3in]{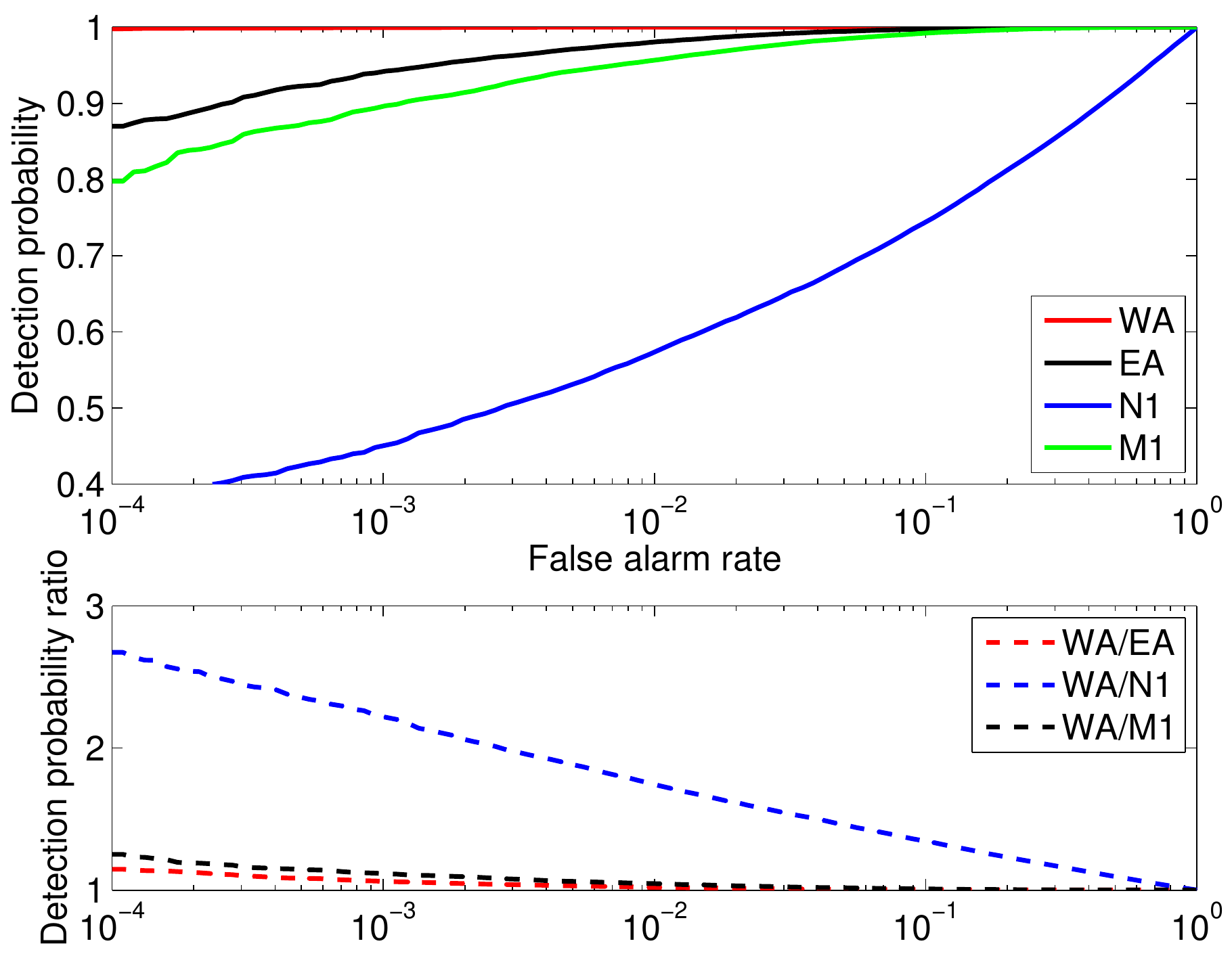}
\includegraphics[angle=0,width=3in]{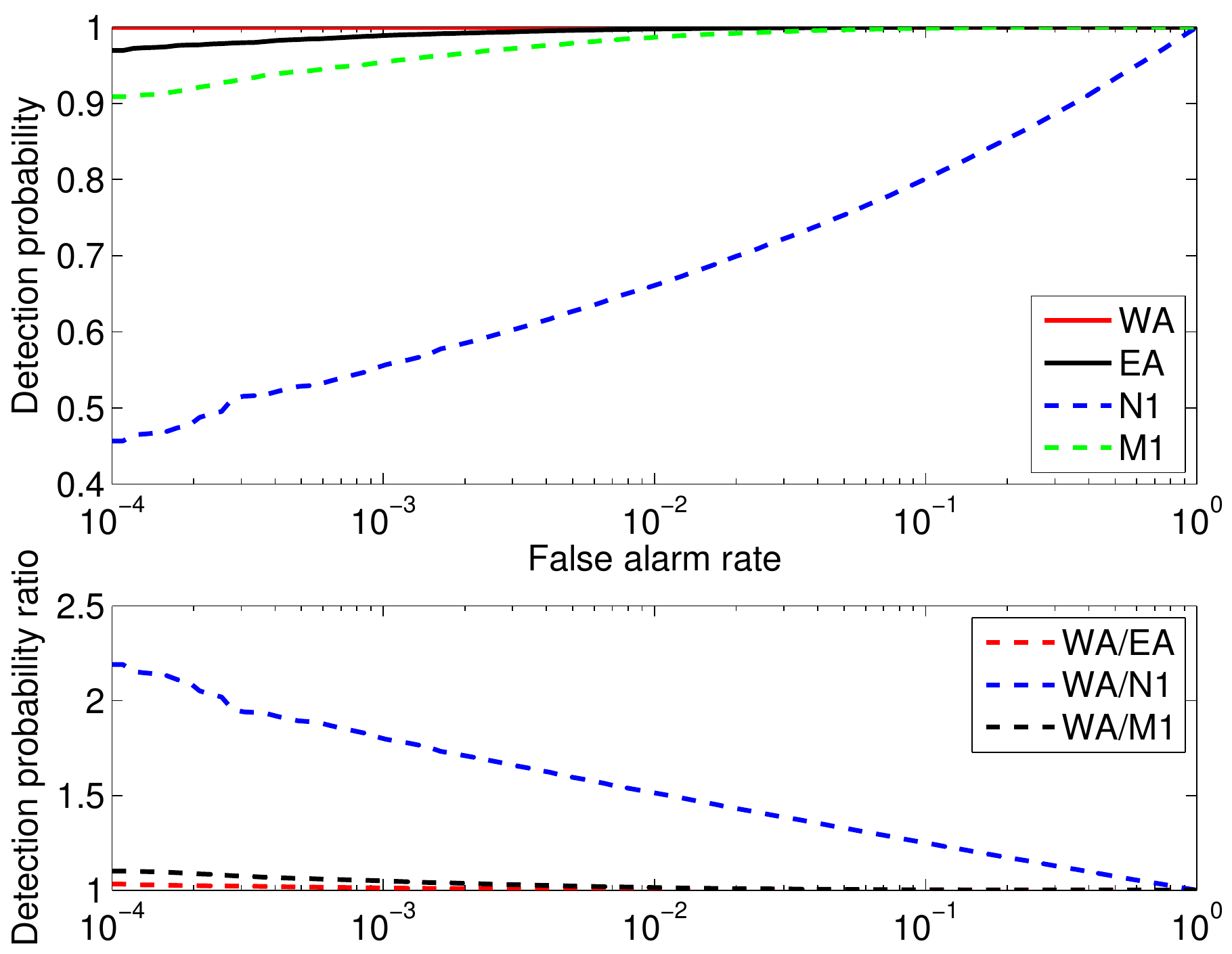}\\
\end{tabular}
    \end{center}
\caption{ROC curves of the weighted-combination (WA), equal-combination (EA),
the expected brightest (N1) and the the measured brightest (M1) detection
methods for $10^5$ signal injection simulations and $10^5$ noise only
simulations. WA and EA correspond to weighted and equal combinations of all
pulsars, respectively.   The intrinsic parameter  in injections was drawn from  two
distributions:  (i) $\epsilon^2$ follows an exponential distribution  with
rate parameter $2\times10^{-16}$  (top left panel) and $4\times10^{-16}$
(bottom-left panel), and (ii) $\epsilon$ follows a Gaussian distribution
with mean value of $1.5\times10^{-8}$ (top-right) and $2\times10^{-8}$
(bottom right) with standard deviations equal to half of the mean value. 
Detection probability ratios (dashed lines) are shown in each panel.}
\label{roc}.
\end{figure*}

As shown in Fig.~\ref{roc}, our proposed weighted-combination method includes
the known information of all sources and detectors, therefore combining all
sources should yield a higher $P_{\rm DE}$ compared to other methods. In the
case of $\overline{\epsilon^2}= 2\times10^{-16}$, given $P_{\rm FA}=0.0001$,
the $P_{\rm DE}{\sim}0.9$ for the weighted-combination method is a factor of
${\sim}2$ to $4$ more sensitive than other methods (see top-left panel of
Fig.~\ref{roc}). The improved performance of the weighted-combination method
over other methods appears to be independent of the ellipticity distribution
types and distribution parameter values used in our simulations.

\begin{figure*}
 \begin{center}
\begin{tabular}{cc}
\includegraphics[angle=0,width=3in]{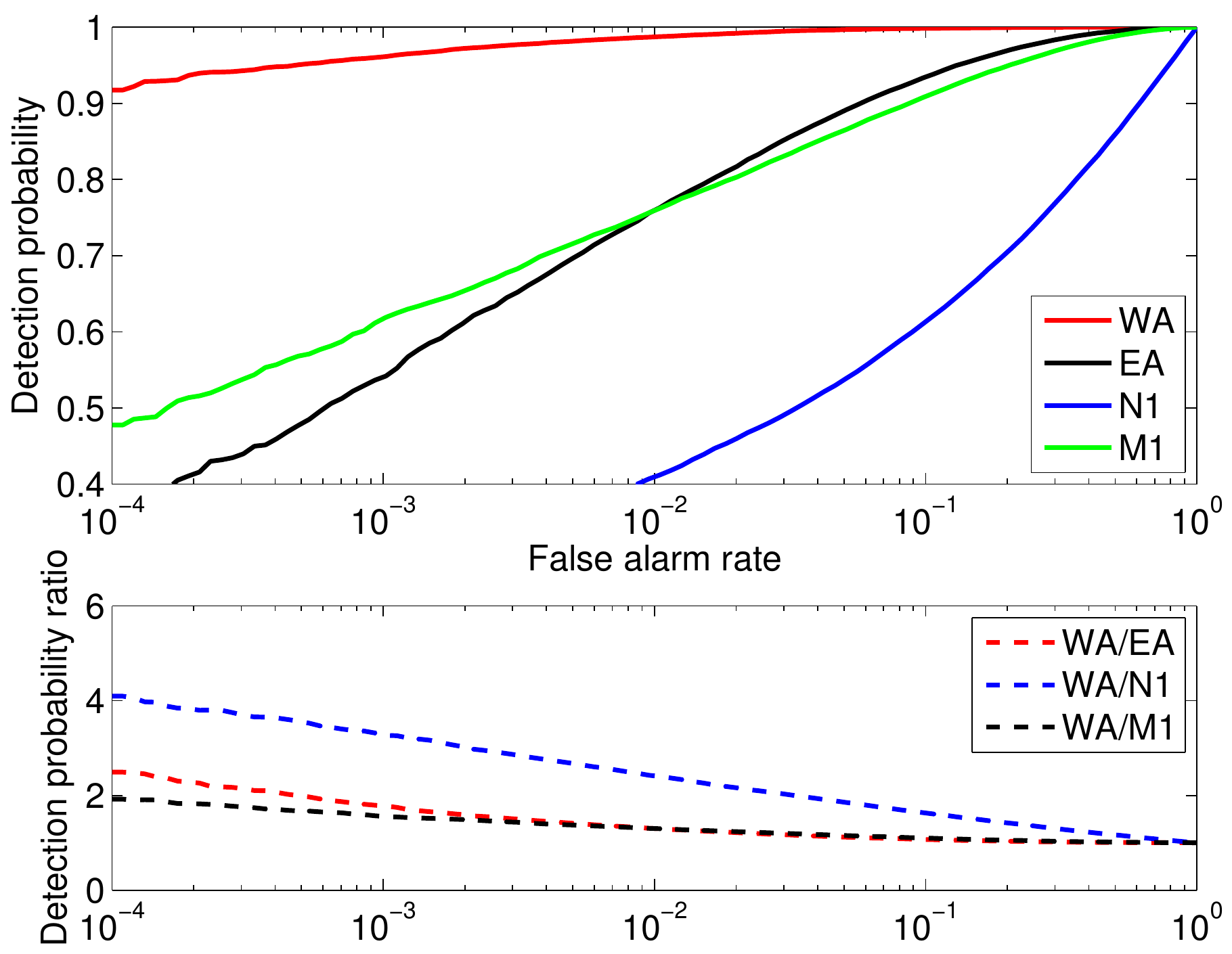} 
\includegraphics[angle=0,width=3in]{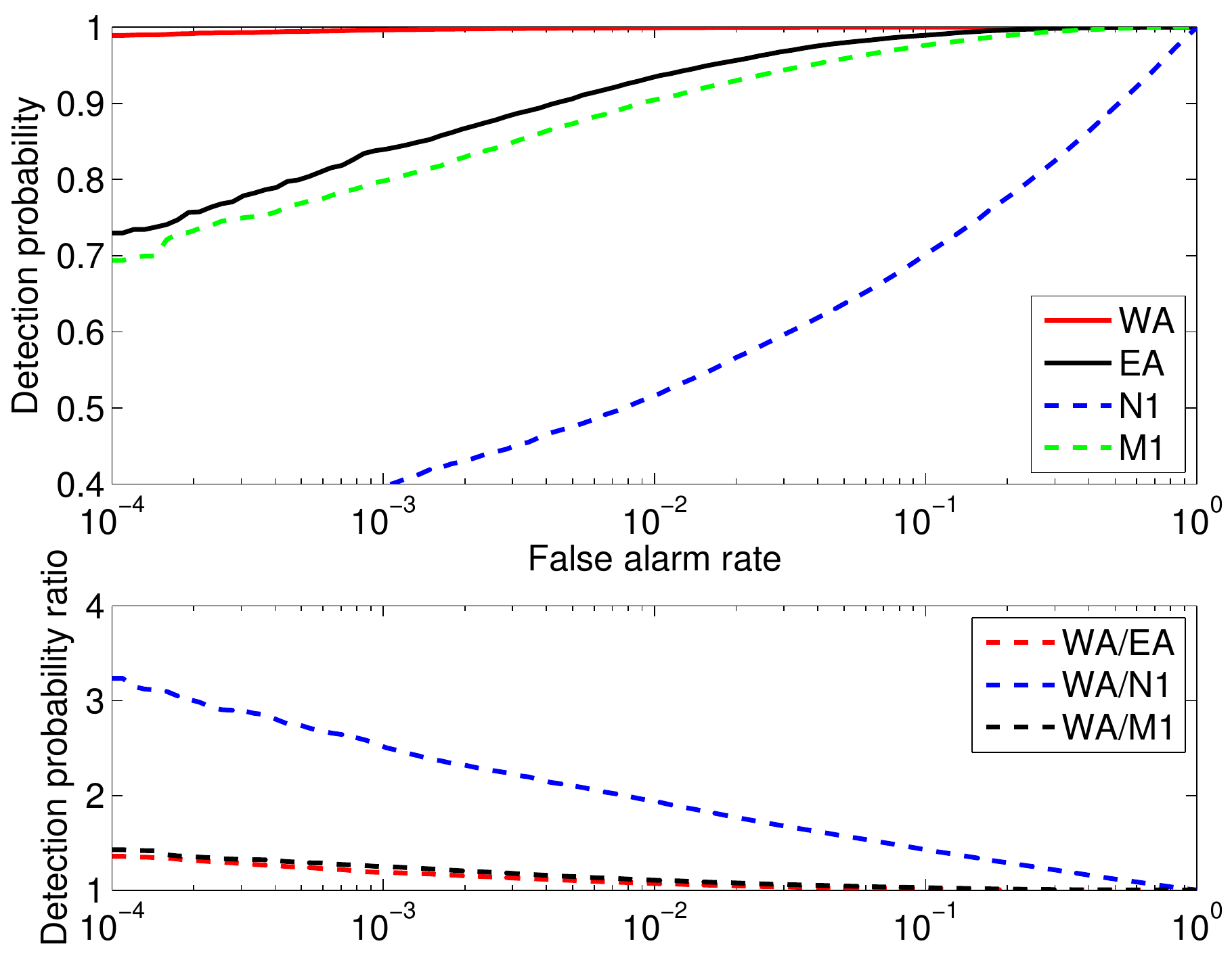}
\\
\end{tabular}
    \end{center}
\caption{The same as for the top panels of Fig.~\ref{roc} but taking into
account the measured distance error. The measured distance error effect is
included by drawing "true" pulsar distances from a normal distribution with
mean value equal to their current best estimates and a standard deviation of
$20\%$ of those mean values.\label{roc_d}}.
\end{figure*}

The typical pulsar distance measurement error is ${\sim}20\%$ but could be up
to a factor of 2-3 larger~\cite[e.g.][]{1993ApJ...411..674T}. To test the
robustness of our proposed method, we test our sensitivity to distance
uncertainty by drawing our pulsar distances from Gaussian distributions with
mean values equal to the best estimated distance and with a standard deviation
equal to $20\%$ of the mean. As shown in Fig.~\ref{roc_d}, the distance
uncertainties do not change the general performance of all methods: our
proposed weighted-combination method still is the most efficient method and
improves the $P_{\rm DE}$ by a factor of $\sim 1.5$ to $4$ compared to
different methods and given $P_{\rm FA}=0.0001$. The level of improvement
decreases when the GW signals become stronger.

 %
\section{Discussion}\label{res}
We have proposed a novel weighted-combination detection statistic for GWs from
an ensemble of known pulsars. The aim of this approach is to improve the
detection efficiency of GWs over that of individual pulsar detection based on
the ${\mathcal F}$-statistic applied to single pulsars. The general argument
behind the combination detection strategy is that a group of sources should be
more detectable than an individual one if they share certain characteristics.
We have shown that our general optimal statistic for the weighted combination
of GW signals outperforms all other approaches.

We have shown that to more efficiently detect GW signals emitted from a
ensemble of pulsars, each source within the ensemble could be assigned a
different detection statistic threshold based on the expected signal strength.
Furthermore, by assuming that the SNRs of all sources are constant or follow
exponential distributions, we have shown that the linearly weighted-combination
statistic is very close to being optimal and is robust to the choice of prior
SNR distributions. These analytic and simple Monte Carlo test predictions are
consistent with results obtained from simulations of known pulsars.

We have also used the ROC function to determine the sensitivity of a range of
possible search strategies where the detection probability between
approaches is compared as a function of false-alarm probability. To
demonstrate the performance of the new weighted-combination detection method
for the Advanced detectors era, we have compared the detection efficiency
of the linearly weighted-combination method versus the equal-combination and individual
detection method. We have done this by simulating GW signals emitted from the 195 known pulsars
within the sensitive frequency band of Advanced LIGO and Virgo. We assume that the
intrinsic pulsar parameter ellipticity $\epsilon^2$ follows a common
distribution in these simulations. The true form of the ellipticity
distribution and its associated parameters are unknown. We have chosen to use
both exponential and Gaussian distributions with mean values corresponding to
ellipticities $\epsilon \sim 10^{-8}$, a value consistent with the initial GW era
nondetection of pulsar signals and a possible advanced era detection. In
general, the combination methods return better detection efficiency than a
method that simply considers the closest or brightest pulsar. Being
consistent with results of simple Monte Carlo tests, the most efficient
method in simulations for known pulsars involves combining all known pulsars
with weights $\propto \overline{\rho}$, the expected value of the optimal SNR
of each pulsar. For the specific case where $\overline{\epsilon} \sim
1.5\times 10^{-8}$, for one year observation of the Advanced detector network,
we find that $P_{\rm DE}{\sim}0.95$ given $P_{\rm FA}=0.0001$. In this case, the
improvement by our proposed combined method could be up to a factor of $\sim 4$
compared with other methods. These results are consistent with the case of
taking into account the measurement errors of pulsar distances.   
  
An important feature of the proposed combination method is that it is very
flexible. Using the new method it is simple to include more observed pulsars
or updated source information (e.g. distance or orientation parameters) ,
without recalculating any individual detection ${\mathcal F}$-statistic
values. However, we would expect that a fully Bayesian approach for combining
all known pulsars may be more sensitive albeit at an increased computational
cost.  

The flagship known pulsar analysis within the GW community is a Bayesian
approach~\cite{2005PhRvD..72j2002D,Pitkin:2011cl,2009CQGra..26t4013P,2014CQGra..31f5002W}.
We note that it is likely that a comprehensive Bayesian approach to combining
all known pulsars into a single analysis may produce a truly optimal result.
Besides all of the  information discussed above, one could also consider the
uncertainty of the major assumption (model) of this work: that all pulsars'
ellipticity values follow a common but unknown distribution.  A hierarchical Bayesian approach would allow us to naturally investigate the true priors governing the distribution.  In this case the form of the prior would be represented as a possible model and the parameters governing that distribution would be the ``hype'' parameters of that model. We could also apply Bayesian model selection to distinguish between different prior distributions e.g. exponential vs Gaussian or power law, etc... However, it is unclear how constraining such an analysis would be and we hope to tackle this problem in future studies.
 Beyond the detection of GWs emitted by a ensemble of pulsars, the posterior probability of
all parameters could be output from a Bayesian approach. In future studies we
hope to investigate such a Bayesian application to the detection of GWs from
the ensemble of known pulsars. 

\acknowledgments
 We would like to acknowledge valuable input from  our anonymous referee, M. Pitkin  and G. Woan,  whose input has greatly
improved the manuscript. XF
acknowledges financial support from National Natural Science Foundation of
China (grant No.~11303009 and 11673008).  XF is a Newton Fellow supported by the Royal
Society. YC is supported by NSF grants PHY-1404569  and C.~M. is
supported by a Glasgow University Lord Kelvin Adam Smith Fellowship and the
Science and Technology Research Council (STFC) grant No. ST/ L000946/1.

\bibliography{Bibliography}
\end{document}